\documentclass{emulateapj}
\usepackage{amstext}
\usepackage{amsmath}

\slugcomment{Accepted for publication in the Astrophysical Journal}
\shorttitle{EVIDENCE FOR NEON-RICH DONORS} 
\shortauthors{JUETT \& CHAKRABARTY}

\begin{document}
\bibliographystyle{apj_noskip}

\title{X-Ray Spectroscopy of the Low-Mass X-ray Binaries 2S~0918--549
and 4U~1543--624: \\Evidence for Neon-Rich Donors}

\author{Adrienne~M.~Juett and Deepto~Chakrabarty\altaffilmark{1}} 
\affil{\footnotesize Department of Physics and Center for Space Research,
	Massachusetts Institute of Technology, Cambridge, MA 02139;\\
	ajuett, deepto@space.mit.edu}

\altaffiltext{1}{Alfred P. Sloan Research Fellow}

\begin{abstract}
We present high-resolution spectroscopy of the neutron-star/low-mass
X-ray binaries 2S~0918$-$549 and 4U~1543$-$624 with the High Energy
Transmission Grating Spectrometer onboard the {\em Chandra X-ray
Observatory} and the Reflection Grating Spectrometer onboard {\em
XMM-Newton}.  Previous low-resolution spectra of both sources showed a
broad line-like feature at 0.7 keV that was originally attributed to
unresolved line emission.  We recently showed that this feature could
also be due to excess neutral Ne absorption, and this is confirmed by
the new high-resolution {\it Chandra\/} spectra.  The {\em Chandra}
spectra are each well fit by an absorbed power-law $+$ blackbody model
with a modified Ne/O number ratio of 0.52$\pm$0.12 for 2S~0918$-$549
and 1.5$\pm$0.3 for 4U~1543$-$624, compared to the interstellar-medium
value of 0.18.  The {\em XMM} spectrum of 2S~0918$-$549 is best fit by
an absorbed power-law model with a Ne/O number ratio of 0.46$\pm$0.03,
consistent with the {\em Chandra} result.  On the other hand, the {\em
XMM} spectrum of 4U~1543$-$624 is softer and less luminous than the
{\em Chandra} spectrum and has a best-fit Ne/O number ratio of
0.54$\pm$0.03.  The difference between the measured abundances and the
expected interstellar ratio, as well as the variation of the column
densities of O and Ne in 4U~1543$-$624, supports the suggestion that
there is absorption local to these binaries.  We propose that the
variations in the O and Ne column densities of 4U~1543$-$624 are
caused by changes in the ionization structure of the local absorbing
material.  It is important to understand the effect of ionization on
the measured absorption columns before the abundance of the local
material can be determined.  This work supports our earlier suggestion
that 2S~0918$-$549 and 4U~1543$-$624 are ultracompact binaries with
Ne-rich companions.
\end{abstract}

\keywords{binaries: close --- 
stars: neutron --- 
stars: individual (2S~0918$-$549) --- 
stars: individual (4U~1543$-$624) --- 
X-rays: binaries}

\section{Introduction}
Low mass X-ray binaries (LMXBs) consist of a neutron star (NS) or
black hole (BH) in orbit with a $\lesssim$1~$M_{\odot}$ companion.  An
intriguing sub-class of LMXBs are the ultracompact binaries that have
orbital periods less than 80 minutes.  Hydrogen-rich companions cannot
sustain a LMXB system with such a short orbital period
\citep*{ps81,rjw82}.  However, orbital periods $\lesssim$80 min are
predicted for hydrogen-deficient or degenerate companions
\citep*{jar78,nrj86}, and this was confirmed with the orbital period
measurements of the X-ray pulsar 4U~1626$-$67 \citep[$P_{\rm orb}$=42
min;][]{mmn+81}, the X-ray dipper 4U~1915$-$05 \citep[$P_{\rm orb}$=50
min;][]{ws82,wbm+82}, the X-ray bursters 4U~1820$-$30 \citep*[$P_{\rm
orb}$=11 min;][]{spw87} and 4U~1850$-$087 \citep[$P_{\rm orb}$=21
min;][]{hcn+96}, and the detection of the white dwarf analogs, the AM
CVn systems \citep[e.g.,][]{w95}.  In addition, the three recently
discovered millisecond X-ray pulsars XTE J1751$-$305, XTE J0929$-$314,
and XTE J1807$-$294 were also found to be ultracompact binaries with
$P_{\rm orb}=$42, 44, and 35~min, respectively
\citep{mss+02,gcm+02,mss03}.  The conventional wisdom has been that
the companions in ultracompact LMXBs are the remains of He white
dwarfs (WDs) that have transferred a significant fraction of their
mass to the NS.  However, recent X-ray spectral evidence indicates
that some companions may be Ne-rich \citep*{scm+01,jpc01}, which
suggests the possibility of C-O or O-Ne-Mg WD companions.  The growing
population of these exotic ultracompact systems, as well as the new
evidence for Ne-rich companions, has interesting implications for the
formation and evolution of binary systems.  Motivated by the
observational evidence for Ne-rich donors, \citet{ynh02} explored
formation scenarios for these systems.  Here we discuss two other
NS/LMXBs that may also be ultracompact binaries.

Both 2S~0918$-$549 ($l=275\fdg9$, $b=-3\fdg8$) and 4U~1543$-$624
($l=321\fdg8$, $b=-6\fdg3$) have been observed by all of the major
X-ray satellites since {\it Uhuru}.  McClintock et al. (1978)
identified the optical counterpart of 4U~1543$-$624 based on the {\it
SAS-3\/} position, which was confirmed by {\it HEAO 1}.  The flux
measurements of 4U~1543$-$624 are roughly constant over the last 25
years \citep{sak94,cs97,adn+00,jpc01,s02,ffm+03} with no periodicities
from 50~s -- 10,000~s found in the {\it EXOSAT\/} data, and no
periodicities from 0.1~s -- 1000~s detected in the {\em SAX} data.
Recently, \citet{s02} presented spectral results from archival {\em
ASCA, SAX}, and {\em RXTE} observations.  These results suggest that
an increase in the luminosity of 4U~1543$-$624 is accompanied by a
hardening of the spectrum.  In addition, an Fe-K emission line is seen
in the hard state, but not detected in the low state.  \citet{ffm+03}
present an independent analysis of the two {\em SAX} observations
which show no significant luminosity change but a spectral hardening
in the second observation.  Both \citet{ffm+03} and \citet{adn+00}
report Fe-K line detections for the {\em SAX} and {\em ASCA}
observations, respectively.

In contrast, 2S~0918$-$549 shows a factor of 10 X-ray variability, but
again has no known periodicities in either the X-ray or optical bands
down to timescales of 1 hour (Forman et al. 1978; Warwick et al. 1981;
Chevalier \& Ilovaisky 1987; Smale \& Lochner 1992; Christian \& Swank
1997; Schulz 1999; Jonker et al. 2001).  Chevalier \& Ilovaisky
(1987), using the {\it Einstein\/} HRI position, identified an
ultraviolet-bright optical counterpart for 2S~0918$-$549 and suggested
a source distance of 15 kpc based on the properties of other LMXBs,
i.e., $M_{V}$$=$0.0 and $(B-V)_{0}$$=$0.0.  Recently, Jonker et
al. (2001) detected a type I X-ray burst from 2S~0918$-$549 and
derived an upper limit to the distance of 4.9 kpc~from radius
expansion arguments.  A similar analysis using {\em BeppoSAX} Wide
Field Camera data of 2S~0918$-$549 implied a distance of 4.2~kpc
\citep{cvz+02}.

We have identified both 2S~0918$-$549 and 4U~1543$-$624 as being part
of a class of four NS/LMXBs all having a similar feature at 0.7~keV in
their low-resolution spectra \citep{jpc01}.  This feature had been
attributed to unresolved line emission from Fe and O \citep*[see,
e.g.,][]{cws94,wka97}.  However, a high-resolution observation of the
brightest of these sources, 4U~0614$+$091, with the {\em Chandra X-Ray
Observatory}, failed to detect any emission lines, finding instead an
unusually high Ne/O number ratio in the absorption along the line of
sight (Paerels et al. 2001).  Previously, we showed that the {\em
ASCA} spectra of all four sources are well fit {\em without} a 0.7~keV
emission line using a model that includes photoelectric absorption due
to excess Ne along the lines of sight and presumably local to the
sources.  In this paper, we present the high-resolution {\em Chandra}
and {\em XMM-Newton} spectra of two sources, 2S~0918$-$549 and
4U~1543$-$624, which support our earlier {\em ASCA} results.

Given the high $L_{\rm X}/L_{\rm opt}$ ratio of these sources and the
excess Ne absorption, we suggested that the systems are ultracompact
binaries with Ne-rich degenerate donors \citep{jpc01}.  This is
similar to the ultracompact X-ray pulsar 4U~1626$-$67.  We note that
while the spectrum of 4U~1626$-$67 does show O and Ne emission lines
\citep{awn+95,scm+01}, it does not resemble the sources we are
considering in either low or high-resolution spectra. While we do
compare our results to those of 4U~1626$-$67, it is important to
consider the differences between the systems when drawing conclusions.

\section{Observations and Data Reduction}
{\it Chandra\/} observed 2S~0918$-$549 on 2000 July 19 and
4U~1543$-$624 on 2000 September 12 for 30 ks each using the High
Energy Transmission Grating Spectrometer (HETGS) and the Advanced CCD
Imaging Spectrometer (ACIS; Weisskopf et al. 2002). The HETGS carries
two transmission gratings: the Medium Energy Gratings (MEGs) with a
range of 2.5--31~\AA\/ (0.4--5.0~keV) and the High Energy Gratings
(HEGs) with a range of 1.2--15~\AA\/ (0.8--10.0~keV).  The HETGS
spectra are imaged by ACIS, an array of six CCD detectors.  The
HETGS/ACIS combination provides both an undispersed (zeroth order)
image and dispersed spectra from the gratings.  The various orders
overlap and are sorted using the intrinsic energy resolution of the
ACIS CCDs.  The first-order MEG (HEG) spectrum has a spectral
resolution of $\Delta\lambda=$ 0.023~\AA\/ (0.012~\AA).

The ``level 1'' event files were processed using the CIAO v2.2 data
analysis package.  For both sources, the standard CIAO spectral
reduction procedure was performed.  During our analysis, it was found
that the pipeline tool {\tt acis\_detect\_afterglow} rejects 3--5\% of
source photons in grating spectra.  Afterglow is the residual charge
left from a cosmic-ray event which is released over several frames and
can cause a line-like feature in a grating spectrum.  The {\tt
acis\_detect\_afterglow} tool flags events that occur at the same chip
coordinates in consecutive frames.  For bright sources, the tool also
rejects a fraction of the source events.  Although only a small
fraction of the total, the rejection of source photons by this tool is
systematic and non-uniform.  Since order-sorting of grating spectra
provides efficient rejection of background events, the afterglow
detection tool is not necessary.  Therefore, we reextracted the event
file, retaining those events tagged by the {\tt
acis\_detect\_afterglow} tool.  No features were found that might be
attributable to afterglow events.

The combined MEG $+$ HEG first order dispersed spectrum of
2S~0918$-$549 has an average count rate of 3.10$\pm$0.18~cts~s$^{-1}$.
We examined the total count rate, as well as the count rates in three
different energy ranges, to check for changes in the spectral state of
2S~0918$-$549.  We found no evidence for any change of state during
the {\it Chandra\/} observation.  For 4U~1543$-$624, the first order
MEG $+$ HEG count rate is 17.2$\pm$0.4~cts~s$^{-1}$.  Similarly, no
changes of state of 4U~1543$-$624 were found.  To look for periodic
modulations of the X-ray flux, we created lightcurves from the
combined first order event files after barycentering and randomizing
the event arrival times.  Randomizing of the event arrival times
consists of adding a random quantity uniformly distributed between
0--3.2~s in order to avoid aliasing caused by the readout time.  We
searched for modulations of the X-ray flux with frequencies between
4$\times10^{-5}$ and 5$\times10^{-2}$ Hz.  We found no evidence for
periodic modulation in either source, with a 90\%-confidence upper
limit of 2\% (2S~0918$-$549) and 1\% (4U~1543$-$624) for the
fractional rms amplitude.

For both observations, we applied the aspect offset correction
available at the {\em Chandra} website\footnote{See
http://asc.harvard.edu/cal/ASPECT/celmon/index.html}.  Using the CIAO
tool {\tt celldetect}, the zeroth order source position for
2S~0918$-$549 was determined: R.A.=$09^{\rm h} 20^{\rm m} 26\fs95$ and
Dec=$-55^{\circ} 12\arcmin 24\farcs7$, equinox J2000.0 (90\%
confidence error of 0\farcs6).  Since the optical counterpart position
is not given by \citet{ci87}, we obtained optical images of the
2S~0918$-$549 field using the Magellan 6.5-m telescopes in Chile.  We
measured the position of the optical counterpart by matching about 20
field stars with the USNO-A2.0 catalog of astrometric standards
\citep{m+98}.  We obtained R.A.=$09^{\rm h} 20^{\rm m} 26\fs99$ and
Dec=$-55^{\circ} 12\arcmin 24\farcs8$ (equinox J2000.0) with an error
radius of 1\farcs8 (90\% confidence).  The optical and X-ray positions
for 2S~0918$-$549 are 1\farcs5 separated.  The zeroth order image of
4U~1543$-$624 was so heavily piled up that no counts were recorded in
the center of the point spread function.  {\tt Celldetect} was not
able to determine an accurate source position, but the CIAO tool {\tt
tgdetect}, which runs celldetect in conjunction with grating specific
filters and sub-tools, was able to position the zeroth order image of
4U~1543$-$624: R.A.=$15^{\rm h} 47^{\rm m} 54\fs69$ and
Dec=$-62^{\circ} 34\arcmin 05\farcs4$, equinox J2000.0.  This position
is $0\farcs2$ from the optical counterpart position given by Bradt \&
McClintock (1983).

For bright sources, pileup can be a problem for CCD detectors
\citep[see, e.g.,][]{d03}. Pileup occurs when two or more photons are
incident on the same pixel during the 3.2~s readout time of the ACIS
detector.  When this happens, the instrument reads just one event at
an energy comparable to the sum of the original photons energies, thus
reducing the detected count rate and modifying the spectral shape.
The zeroth order {\it Chandra\/} spectra of both 2S~0918$-$549 and
4U~1543$-$624 were heavily affected by pileup (pileup fraction
$>$80\%) and were not used in this analysis.  In addition, pileup can
also affect dispersed spectra.  We checked the dispersed spectra of
both sources and found that for 2S~0918$-$549 no pileup occurred;
whereas for 4U~1543$-$624, there were signs of pileup in the first
order MEG spectrum.  In dispersed spectra, pileup can affect only a
limited wavelength range, particularly where the effective area of the
instrument is the highest.  For 4U~1543$-$624, pileup occurred in the
range 5--12~\AA\/ (1--2.5~keV).

For 2S~0918$-$549, we used the standard CIAO tools to create detector
response files (ARFs) for the MEG and HEG $+1$ and $-1$ order spectra.
These were combined when the $+/-$ order spectra were added for the
MEG and HEG separately.  The data were binned to 0.03~\AA\/
(0.015~\AA) for the MEG (HEG) to ensure good statistics.  We also
created background files for the MEG and HEG spectra using the
standard CIAO tool.

There are known differences in the quantum efficiencies (QEs) of the
front-illuminated (FI) CCDs compared to the back-illuminated (BI)
CCDs.  An initial analysis using the standard QE files revealed that,
for 4U~1543$-$624, the chip differences were identifiable (especially
at high wavelengths) and needed to be corrected.  Using a correction
table\footnote{Documentation and table available at
http://space.mit.edu/CXC/calib/hetgcal.html} provided by
H.~L. Marshall of the HETGS instrument team, we modified the QE files
available in the standard CIAO calibration database and then used them
to create ARFs for the spectral analysis of 4U~1543$-$624.  This
modification was not required for the 2S~0918$-$549 data.  In order to
use the grating pileup kernel in ISIS, we created ARFs for each chip
using the CIAO tool {\tt mkgarf} and then combined them to create $+1$
and $-1$ order MEG and HEG ARFs using a custom tool developed by
J. Davis of the HETGS instrument team.  (This tool is similar to the
standard CIAO tool but also correctly calculates the fractional
exposure at each response bin.  This functionality has been added to
the standard tools in CIAO v2.3.)  The grating pileup kernel models
the effect of pileup on grating spectra, similar to the pileup model
available for CCD spectra in ISIS and XSPEC \citep[see,][for a
discussion of grating pileup modeling]{d03}.  In addition, the data
and responses were rebinned to 0.033~\AA\/ (0.017~\AA) for the MEG
(HEG) to reflect the size of an ACIS event detection cell (3 CCD
pixels).  We used this binning for the spectral analysis.  All {\em
Chandra} spectral analysis of 2S~0918$-$549 and 4U~1543$-$624 was
performed using ISIS \citep{hd00}.

{\em XMM} observed 2S~0918$-$549 on 2001 May 5 for 40 ks and
4U~1543$-$624 on 2001 Feb 4 for 50 ks.  {\em XMM} carries three
different instruments, the European Photon Imaging Cameras
\citep[EPICs;][]{sbd+01,taa+01} the Reflection Grating Spectrometers
\citep[RGSs;][]{hbk+01}, and the Optical Monitor
\citep[OM;][]{mbm+01}.  The OM was not used in this analysis.

The EPIC instruments consist of 3 CCD cameras, MOS 1, MOS 2, and pn,
which provide imaging, spectral and timing data.  For both
observations, the pn and MOS 1 cameras were run in timing mode which
provides high time resolution event information by sacrificing
1-dimension in positional information.  During the 2S~0918$-$549
observation, the MOS 2 camera was run in full-frame mode, while for
the 4U~1543$-$624 observation, it was run in small-window mode.  The
EPIC instruments provide good spectral resolution ($\Delta
E$$=$50--200~eV FWHM) over the 0.3--12.0 keV range.  There are two
RGSs onboard {\em XMM} which provide high-resolution spectra
($\lambda/\Delta\lambda$$=$100--500 FWHM) over the 5--38 \AA\/ range.
The grating spectra are imaged onto CCD cameras similar to the
EPIC-MOS cameras which allows for order sorting of the high-resolution
spectra.

The {\em XMM} data were reduced using SAS version 5.3.  Standard
filters were applied to all {\em XMM} data and times of high
background were excluded.  The EPIC-pn data were reduced using the
procedure {\tt epchain}.  Source and background spectra were extracted
using filters in the one spatial coordinate, DETX.  A ready-made
response file for pn timing mode data can be found on the {\em XMM}
website\footnote{Available at http://xmm.vilspa.esa.es/ccf/epic/}.  We
reduced the EPIC-MOS data using {\tt emchain}.  Since no response file
was available for MOS timing mode observations, the MOS 1 data were
not used in this analysis.  The MOS 2 data were found to have
considerable pileup.  To reduce the effect of pileup on the spectra,
we extracted source spectra using an annulus to excise the core of the
point spread function where pileup is prevalent.  The inner
20\arcsec\/ were excluded for 2S~0918$-$549 and the inner 35\arcsec\/
for 4U~1543$-$624.  Background spectra were extracted using an annulus
centered on the source but excluding any other point sources in the
field of view.  Response files were generated for the MOS 2 spectra
using {\tt rmfgen} and {\tt arfgen}.  The RGS data were reduced using
{\tt rgsproc}, which produced standard first order source and
background spectra and response files for both RGSs.  We grouped the
pn and MOS spectra to oversample the energy resolution of the CCDs by
no more than a factor of three.  The RGS spectra were grouped to
ensure that each bin had at least 20 counts.  All {\em XMM} spectral
analysis of 2S~0918$-$549 and 4U~1543$-$624 was performed using XSPEC
\citep{a96}.

The 2S~0918$-$549 observation had average count rates of
81.06$\pm$0.05, 13.03$\pm$0.02, and 8.928$\pm$0.016 cts s$^{-1}$ for
the pn, MOS 2, and combined RGS, respectively.  Using the pn data, we
checked for changes in the spectral state by examining the total count
rate, as well as the count rates in three different energy ranges.
There was no evidence for a change in the spectral state of
2S~0918$-$549 during the {\em XMM} observation.  The average count
rates during the 4U~1543$-$624 observation were 238.89$\pm$0.07,
46.88$\pm$0.03, and 23.07$\pm$0.02 cts s$^{-1}$ for the pn, MOS 2, and
combined RGS, respectively.  Again, no change in the spectral state of
4U~1543$-$624 was found during the {\em XMM} observation.  While the
pn timing mode allows for a time resolution of 0.03 ms, searches for
periodic and quasi-periodic modulation at high frequency are hampered
by a 167 Hz instrumental signal (and multiple harmonics).  Due to the
low count rates of these observations ($\ll$1 count per readout), the
low frequency end of the power spectrum is affected by instrumental
noise \citep[see,][and references therein]{kkb+02}.  To remove the
noise, we calculated the average power at each frequency using an 11
bin moving average and then divided by this value.  We verified that
the powers were distributed correctly and then searched for modulation
of the X-ray flux with frequencies between $2\times10^{-5}$ and 0.5
Hz.  No modulation was detected with a 90\%-confidence upper limit on
the fractional rms amplitude of 0.3\% (2S~0918$-$549) and 0.14\%
(4U~1543$-$624).

\section{Spectral results}
\subsection{2S~0918$-$549}
To determine the appropriate continuum model, we fit the {\em Chandra}
HEG and MEG spectra jointly with both an absorbed power-law and an
absorbed power-law $+$ blackbody model over the wavelength ranges
1.5--13~\AA\/ for the HEG and 1.8--25~\AA\/ for the MEG.  In all
cases, we used the interstellar absorption model of \citet*[{\tt
tbabs};][]{wam00}.  In addition, we included an edge at 43.7~\AA\/ to
account for the instrumental C buildup on the ACIS CCDs\footnote{For
more information see,
http://cxc.harvard.edu/cal/Links/\\Acis/acis/Cal\_prods/qeDeg/index.html}
\citep{psm+02}.  We calculated that the appropriate optical depth for
the instrumental C edge at the time of the 2S~0918$-$549 observation
was 0.79 from calibration data points provided by H.~L.~Marshall.
This edge is included in all {\em Chandra} spectral fits of
2S~0918$-$549.  The power-law $+$ blackbody model produced a
significantly better chi-squared value than the power-law alone
(significance greater than 99.99\% as calculated by an F-test), so we
take a power-law $+$ blackbody model as the appropriate continuum
model for the {\em Chandra} spectra.  The best-fit parameter values
are given in Table 1.

\begin{deluxetable*}{lcccccc}
\tablecaption{Spectral Fits\tablenotemark{a}} 
\tablehead{ & \multicolumn{1}{c}{Absorption} & \multicolumn{2}{c}{Power-law} &
\multicolumn{2}{c}{Blackbody} &  \\ 
\cline{1-2} \cline{3-4} \cline{5-6} & \colhead{$N_{\rm H}$} & & & 
\colhead{$kT$} & & \\ 
\colhead{Model\tablenotemark{b}} & \colhead{(10$^{21}$ cm$^{-2}$)} & 
\colhead{$\Gamma$} & \colhead{$A_{1}$\tablenotemark{c}} & \colhead{(keV)} & 
\colhead{$R^2_{\rm km}/d^2_{\rm 10 kpc}$} & \colhead{$\chi^2$/dof}}
\startdata
\multicolumn{7}{c}{2S~0918$-$549} \\ \tableline
{\em Chandra} tbabs$\times$(PL+BB) & 2.4$\pm$0.3 & 2.04$\pm$0.08 & 2.9$\pm$0.3 
& 0.54$\pm$0.02 & 30$\pm$9 & 1.11 \\
{\em Chandra} full model & 2.9$\pm$0.3 & 2.10$\pm$0.09 & 3.3$\pm$0.5 & 
0.55$\pm$0.04 & 23$\pm$9 & 1.07 \\ \tableline
{\em XMM} tbabs$\times$PL & 2.95$\pm$0.02 & 2.265$\pm$0.004 & 6.23$\pm$0.03 & 
\nodata & \nodata & 1.61 \\
{\em XMM} full model & 3.20$\pm$0.03 & 2.25(fixed) & 5.48$\pm$0.06 & \nodata & 
\nodata & 1.27 \\ \tableline
\multicolumn{7}{c}{4U~1543$-$624} \\ \tableline 
{\em Chandra} tbabs$\times$(PL+BB) & 1.58$\pm$0.11 & 
1.539$^{+0.03}_{-0.004}$ & 12.6$\pm$0.6 & 0.62$\pm$0.03 & 61$\pm$13 & 1.75 \\
{\em Chandra} full model & 3.31$\pm$0.014 & 1.97(fixed) & 17.1$\pm$0.8 & 
\nodata & \nodata & 1.32 \\ \tableline
{\em XMM} tbabs$\times$(PL+BB) & 3.478$\pm$0.017 & 2.568$\pm$0.008 & 
25.12$\pm$0.11 & 1.480$\pm$0.005 & 3.38$\pm$0.07 & 4.34 \\
{\em XMM} full model & 3.613$\pm$0.014 & 2.7(fixed) & 16.26$\pm$0.06 & 
1.512(fixed) &  18.5$\pm$0.3 & 1.87 
\enddata
\tablenotetext{a}{All errors are quoted at the 90\%-confidence level}
\tablenotetext{b}{PL=power law; BB=blackbody; tbabs=photoelectric absorption 
model of \citet{wam00}; full model=absorption of O, Fe, and Ne modeled by edge
models with O absorption lines where needed.  The results for the edge models 
are given in Table 2.}  
\tablenotetext{c}{Power-law normalization at 1 keV in units of 10$^{-2}$ 
photons~keV$^{-1}$ cm$^{-2}$ s$^{-1}$}
\end{deluxetable*}

We showed that the lower-resolution {\em ASCA} spectra of
2S~0918$-$549 could be well fit by including photoelectric absorption
with non-solar abundances of O and Ne \citep{jpc01}.  High-resolution
spectroscopy can test the accuracy of this hypothesis by directly
measuring the edge depths, and therefore column densities, of neutral
O, Fe, and Ne.  We can then compare the abundance ratios to those
expected for the interstellar medium (ISM).  To determine the columns
of each of these elements, we replaced the absorption model with the
comparable variable abundance absorption model \citep[{\tt
tbvarabs};][]{wam00} and set the abundances of O, Fe and Ne to zero.
The $N_{\rm H}$ parameter was allowed to vary while all other
absorption model parameters were frozen.  The Ne and O-K edges were
each modeled by a multiplicative edge model of the form:
\begin{equation}
M(E)=
\begin{cases}
1 & \text{for $E< E_{\rm edge}$}, \\
\exp\:[-\tau (E/E_{\rm edge})^{-3}] & \text{for $E\geq E_{\rm edge}$},
\end{cases}
\end{equation}
with $\tau$ and $E_{\rm edge}$ as variable parameters.  Individual
edge models are used since the edge positions and shapes in the
standard absorption models are not appropriate for high-resolution
spectral analysis (see, Wilms et al. 2000; Juett et al. 2003, in
preparation).  A Gaussian line was included to model the interstellar
absorption line due to atomic O at 23.5~\AA.  The more complex Fe-L
edge at 17.5~\AA\/ was modeled using a table model, created from the
optical constant measurements by Kortright \& Kim (2000), which
allowed for both the column density of Fe and the edge energy to vary.
The continuum was again described by a power law $+$ blackbody.  The
resulting parameters for the power-law and blackbody models are
consistent with the low-resolution fit.  The results of the fit can be
found in Table~1 and are shown in Figure 1.

\begin{figure}
\epsscale{1.2}
\plotone{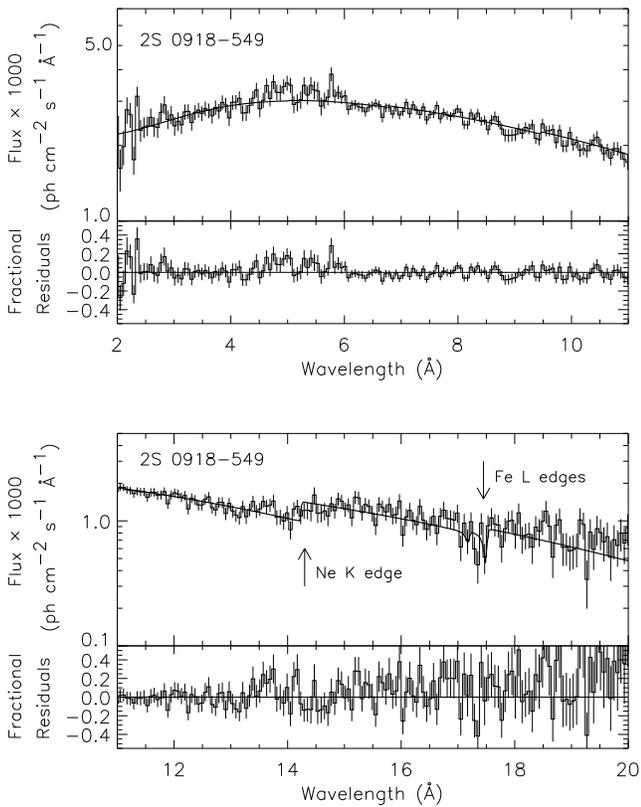}
\caption{(upper panels) {\it Chandra\/} MEG first order spectrum of
2S~0918$-$549 with best-fit power-law $+$ blackbody model including
the O, Fe, and Ne edge models.  For clarity, the data is binned to
0.06~\AA.  (lower panels) Fractional residuals ([data$-$model]/model)
of the MEG first order spectral fit shown above.  There are no
features detected which are consistent with narrow emission or
absorption lines.  The arrows mark the Ne-K and Fe-L absorption edges
at 14.3 and 17.5~\AA\/ respectively.}
\label{fig:1} 
\end{figure}

From the absorption edge depths, the absorbing column densities of O,
Fe, and Ne can be determined along with the equivalent hydrogen column
for each element assuming interstellar abundances.  These results are
summarized in Table 2.  The atomic cross-sections of \citet*{hgd93}
were used for O and Ne, while the Fe-L cross-section was taken from
\citet{kk00}.  We use the ISM abundances given by \citet{wam00} to
calculate equivalent hydrogen column densities.  The wavelengths of
the edges are consistent with measurements of neutral ISM edges in
other LMXBs (Juett et al. 2003).  We find a Ne column density of
$N_{\rm Ne}=(9.9\pm1.7)\times10^{17}$~cm$^{-2}$ and an O column
density of $N_{\rm O}=(1.9\pm0.3)\times10^{18}$~cm$^{-2}$ (all errors
quoted at 90\%-confidence level).  The equivalent hydrogen column
implied by the Ne edge is roughly three times that calculated from the
O edge.  We find a Ne/O number ratio of 0.52$\pm$0.12, compared to the
ISM ratio of 0.18.  The measured Fe column is $N_{\rm
Fe}=(10^{+6}_{-3})\times10^{16}$ cm$^{-2}$.  The atomic O absorption
line has a wavelength of 23.52$\pm$0.04~\AA\/ and an equivalent width
(EW) of 0.017$\pm$0.010~\AA.

The evidence for Ne enhancement in the absorption to 2S~0918$-$549 is
similar to what is seen in 4U~1626$-$67 \citep{scm+01}.  Since the
spectrum of 4U~1626$-$67 also shows emission lines from Ne and O, we
performed a careful search of the {\em Chandra} spectral residuals of
2S~0918$-$549 to place limits on the presence of any spectral line
features.  Gaussian models with fixed FWHM$=$2000 km s$^{-1}$ were fit
at each point in the wavelength range 1.5--13~\AA\/ for the HEG and
1.8--25~\AA\/ for the MEG.  From this, we can place a 3$\sigma$ upper
limit on the EW of any line feature, either emission or absorption.
The EW limit increases with wavelength, varying from 0.02~\AA\/ at
1.5~\AA\/ to 0.5~\AA\/ at 23~\AA.  These EW limits are 3--10 times
lower than the line EWs found in the {\em Chandra} observation of
4U~1626$-$67.  Since the lines found in the spectrum of 4U~1626$-$67
were broader, FWHM of 5340 and 7460 km s$^{-1}$ for the \ion{Ne}{10}
and \ion{O}{8} lines fit by single Gaussians, we investigated how the
EW limit changes as the model FWHM is increased.  Increasing the FWHM
of the line to 5000 (7000) km s$^{-1}$, increases the EW upper limits
by a factor of 1.5 (2.0).

To determine the best-fit continuum model for the {\em XMM} spectra of
2S~0918$-$549, we simultaneously fit the pn, MOS 2, and both RGS
spectra.  Since the pn timing mode response is not well calibrated
below 1 keV, we only fit the pn spectrum in the 1.0--12.0 keV range.
The energy range used for the MOS 2 was 0.3--8.0~keV and for the two
RGS spectra it was 0.35--2.0~keV.  We again fit both an absorbed
power-law and an absorbed power-law $+$ blackbody model.  We also
included a multiplicative constant to account for any instrumental
normalization differences.  During the {\em XMM} observation, the
blackbody component was not significantly detected and we choose an
absorbed power-law as the appropriate continuum model.  The best-fit
parameter values for the {\em XMM} spectral fit are given in Table 1.

For the high-resolution measurements of the absorption edge depths, we
fit only the 2 RGS spectra using a power-law as the continuum model.
Since the RGS covers only a limited wavelength range, 6.2--37.6~\AA,
we froze the power-law index to the best-fit value of the continuum
model fit.  Again we used the variable absorption model with O, Fe and
Ne abundances set to zero, and included the appropriate edge models
for O, Fe, and Ne.  A Gaussian model was used to model the atomic O
absorption line.  The best-fit results are given in Tables 1 and 2 and
shown in Figure 2.  From the fits of the {\em XMM} RGS spectra, we
find a Ne column density of $N_{\rm Ne}=(9.1\pm0.6)\times10^{17}$
cm$^{-2}$ and an O column density of $N_{\rm
O}=(1.96\pm0.07)\times10^{18}$ cm$^{-2}$, both consistent with the
{\em Chandra} measurements.  We find a Ne/O number ratio of
0.46$\pm$0.03.  The best-fit atomic O absorption line had a wavelength
of 23.504$\pm$0.013~\AA\/ and an EW of 0.020$\pm$0.005~\AA.  The
initial best-fit Fe-L column density left large residuals around the
edge shape suggesting that the value was too large.  It is possible
that fits of the entire RGS wavelength range are not as sensitive to
the Fe-L complex.  To determine the correct Fe column, we fit only the
16.5--18~\AA\/ range with a simple model consisting of a power-law and
the Fe-L table model.  The power-law index was allowed to vary for
this fit.  From this fit, we found a Fe column of
$(8\pm3)\times10^{16}$ cm$^{-2}$, which was a much better visual match
and consistent with the {\em Chandra} value.  Our values for other
parameters were determined with the Fe column density fixed to the
above value.

We also searched for line features in the {\em XMM} spectrum of
2S~0918$-$549 using the same procedure as for the {\em Chandra}
spectra.  Gaussian models with fixed FWHM$=$2000 km s$^{-1}$ were fit
at each point in the wavelength range 6.25--37.6~\AA\/ for both RGS
spectra.  We place a 3$\sigma$ upper limit of 0.05~\AA\/ on the EW of
any line feature in the {\em XMM} spectrum of 2S~0918$-$549 in the
wavelength range 6--18~\AA.  At larger wavelengths, the EW limit
increases to 0.6~\AA\/ at 35~\AA.  Again, these EW limits are much
lower than found in the 4U~1626$-$67 spectrum.  If the FWHM of the
Gaussian is increased to 5000~km~s$^{-1}$, then the EW limit increases
by roughly a factor of 2.  There are residuals in the fit around
14~\AA\/ and 19~\AA\/ at the 20\% level (see, Figure 2).  The
residuals of the 4U~1543$-$624 RGS spectral fit are similar (see,
Figure 5), while the {\em Chandra} fit of 2S~0918$-$549 does not show
the same features (see, Figure 1).  This suggests that the residuals
are due to remaining uncertainty in the instrument calibration.

\begin{figure}
\epsscale{1.2}
\plotone{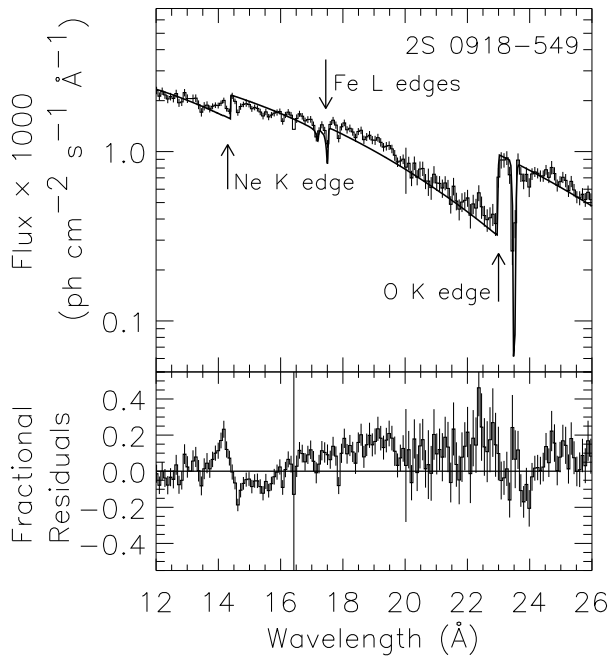}
\caption{(upper panel) {\em XMM} combined RGS first order spectrum of
2S~0918$-$549 with best-fit power-law model including the O, Fe, and
Ne edge models.  For clarity, the data is binned to 0.09~\AA.  (lower
panel) Fractional residuals ([data$-$model]/model) of the RGS first
order spectral fit shown above.  The absorption line at 23.5~\AA\/ is
the expected atomic O line.  There are no other features detected
which are consistent with narrow emission or absorption lines.  The
arrows mark the Ne-K, Fe-L and O-K absorption edges at 14.3, 17.5, and
23~\AA\/ respectively.}
\label{fig:2} 
\end{figure}

\begin{figure}
\plotone{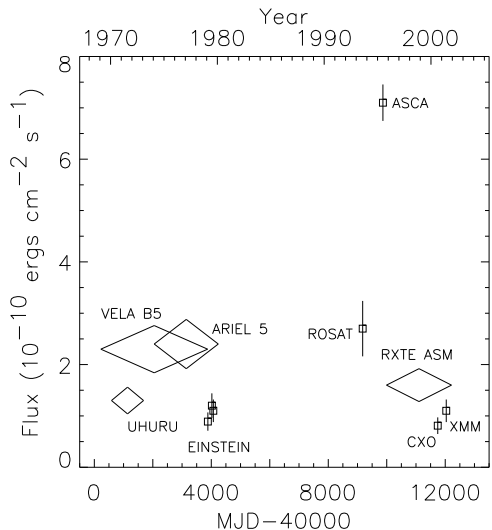}
\caption{The 2--10 keV absorbed flux of 2S~0918$-$549 in units of
$10^{-10}$~ergs~cm$^{-2}$~s$^{-1}$ over the past 30 years. The
majority of the observations are reasonably consistent, except for the
1995 {\it ASCA\/} observation, when the source seems to have been in a
bright state.  The large diamonds represent average flux measurements
for monitoring missions, while the small squares represent pointed
observations.}
\label{fig:3}
\end{figure}

The absorbed 2--10 keV flux of 2S~0918$-$549 was $8.1\times10^{-11}$
and $1.1\times10^{-10}$ erg~cm$^{-2}$~s$^{-1}$ for the {\em Chandra}
and {\em XMM} observations, respectively.  This is an order of
magnitude lower than the {\em ASCA} observation in 1995.  The {\em
Chandra} and {\em XMM} flux measurements are, however, consistent with
other past observations, including the {\em RXTE} ASM flux
measurements made contemporaneously (see, Figure 3).  The spectral
shape of the source has been roughly constant over the past 20 years,
excluding the addition of a blackbody component in some more recent
data.  Although there is some difference in the best-fit slope of the
power-law in each observation ($\Gamma=1.73$ for {\em ASCA}, 2.04 for
{\em Chandra}, and 2.265 for {\em XMM}), this difference is not enough
to account for the observed change in flux.  Rather, the source seems
to be brighter during the {\em ASCA} observation with larger
normalizations measured for both the power-law and blackbody
components.  We suggest that 2S~0918$-$549 was in a bright state
during the {\em ASCA} observation, similar to outbursts seen in other
LMXBs \citep[e.g.,][]{hk00}.  As seen in X-ray color-color diagrams
for sources in outburst, the hard color varies between outburst and
quiescent phases while the soft color remains roughly the same
\citep[see,][]{mrc02}.  {\em ASCA, Chandra} and {\em XMM} have energy
sensitivity which cover only the soft color energy range in standard
color-color diagrams, so a spectral difference during an outburst may
not be measurable with these instruments.

\subsection{4U~1543$-$624}
We fit the four ($+1$ and $-1$; MEG and HEG) {\em Chandra} spectra of
4U~1543$-$624 jointly with both an absorbed power-law and an absorbed
power-law $+$ blackbody model in the wavelength range 1.7--25~\AA\/
for the MEG and 1.3--13~\AA\/ for the HEG.  The pileup kernel in ISIS
was used to model the pileup in the grating spectra.  In addition, an
edge model at 43.7~\AA\/ was included to account for the instrumental
C edge.  We calculated that the appropriate optical depth for the
instrumental C edge at the time of the 4U~1543$-$624 observation was
0.85.  This edge is included in all {\em Chandra} spectral fits of
4U~1543$-$624.  The absorbed power-law $+$ blackbody model was a
significantly better fit to the data and we take this as the
appropriate continuum model.  The absorbed 2--10 keV flux of
4U~1543$-$624 during the {\em Chandra} observation was 7.1$\times
10^{-10}$ erg~cm$^{-2}$~s$^{-1}$, which is consistent with past
observations.

As was done with the 2S~0918$-$549 observations, we modeled the O and
Ne edges with edge models and the Fe-L edge with the table model based
on the latest cross-section from \citet{kk00}.  A Gaussian model was
included to fit the atomic O absorption line at 23.5~\AA.  Since the
most interesting features in the spectrum are at $>12$~\AA, where
pileup is no longer a problem, we fit the MEG spectra in the
wavelength range 12--25~\AA\/ without the pileup kernel to measure the
edge features.  Pileup in the grating spectra enhances the
instrumental residuals found at 2~keV, inflating the chi-squared value
of the fit.  By fitting only the high wavelength end of the spectra,
we determine a more characteristic chi-squared value which can be used
to judge the goodness of the fit.  The blackbody component is only
prominent at low wavelengths, so we chose to model the continuum in
this limited wavelength range as a power-law.  We found that if the
equivalent hydrogen column density of the absorption model and the
power-law index are allowed to vary, there is a degeneracy between the
parameters that drives $N_{\rm H}$ and $\Gamma$ to higher values.  To
remove this degeneracy, we fixed the photon index to that found in a
continuum fit with the full model.  The results can be found in Table
2.

From the absorption edge depths, we find a Ne column density of
$N_{\rm Ne}=(12.4\pm1.1)\times10^{17}$ cm$^{-2}$ and an O column
density of $N_{\rm O}=(8.5\pm1.4)\times10^{17}$ cm$^{-2}$.  The
equivalent hydrogen column implied by the Ne edge is more than eight
times that calculated from the O edge.  We find a Ne/O number ratio of
1.5$\pm$0.3, compared to the ISM ratio of 0.18.  The atomic O
absorption line was found at 23.504$\pm$0.015~\AA\/ with an EW of
0.04$\pm$0.02~\AA.  The measured Fe column is $N_{\rm
Fe}=(3\pm2)\times10^{16}$ cm$^{-2}$.

As can be seen in Figure 4, no emission or absorption lines are
apparent in the spectrum of 4U~1543$-$624.  We performed a careful
search of the {\em Chandra} spectral residuals to place limits on the
presence of any spectral features in the 12--25~\AA\/ wavelength range
of the MEG.  Gaussian models with fixed FWHM$=$2000 km s$^{-1}$ were
fit at each point.  From this, we can place a 3$\sigma$ upper limit on
the EW of any line feature, either emission or absorption.  The EW
limit increases with wavelength, varying from 0.02~\AA\/ at 12~\AA\/
to 1.0~\AA\/ at 23~\AA, while broader line models give EW upper limits
a factor of 2 larger.  The EW limits found for the {\em Chandra}
spectrum of 4U~1543$-$624 are lower than the emission line EWs found
in the spectrum of 4U~1626$-$67.  There have been a number of reports
of a Fe K line detection in the spectrum of 4U~1543$-$624.  There is
no apparent line in the Fe K emission region (6.4--6.7~keV) of the
{\em Chandra} spectrum.  To place a limit on the EW of such a line, we
fit an absorbed power-law $+$ blackbody $+$ Gaussian model to the full
spectrum with a fixed line energy of 6.4~keV.  The upper limit on the
EW of the line is 60 eV.

\begin{figure}
\epsscale{1.2}
\plotone{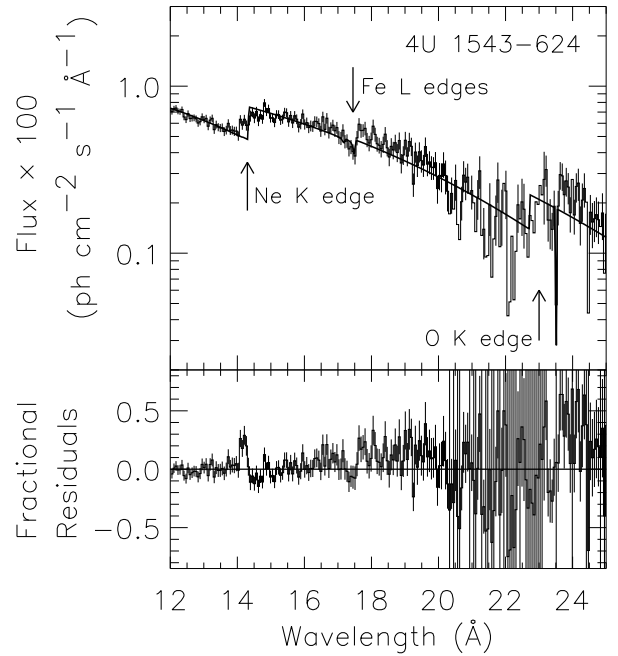}
\caption{(upper panel) {\em Chandra} combined MEG first order spectrum
of 4U~1543$-$624 with best-fit power-law model including the O, Fe,
and Ne edge models.  For clarity, the data is binned to 0.06~\AA.
(lower panel) Fractional residuals ([data$-$model]/model) of the MEG
first order spectral fit shown above.  The absorption line at
23.5~\AA\/ is the expected atomic O line.  There are no other features
detected which are consistent with narrow emission or absorption
lines.  The arrows mark the Ne-K, Fe-L and O-K absorption edges at
14.3, 17.5, and 23~\AA\/ respectively.}
\label{fig:4} 
\end{figure}

The analysis of the {\em XMM} spectra of 4U~1543$-$624 is similar to
the {\em XMM} analysis of 2S~0918$-$549.  We first jointly fit the
EPIC-pn, MOS 2, and RGS spectra with both an absorbed power-law and an
absorbed power-law $+$ blackbody.  For 4U~1543$-$624, the power-law
$+$ blackbody was the best-fit continuum model.  During the {\em XMM}
observation, the absorbed 2--10~keV flux was 4.2$\times 10^{-10}$
erg~cm$^{-2}$~s$^{-1}$ which at the lower end of the range of observed
flux of the source.  The best-fit parameters of the {\em XMM} spectral
fit of 4U~1543$-$624 has a steeper power-law photon index than found
in the {\em Chandra} observation of the source.  In addition, the
blackbody temperature and normalization are different, although this
difference could be due to incorrectly accounting for the absorption
features.  If the full {\em Chandra} spectrum is fit with a model that
allows for a non-ISM Ne/O ratio, the blackbody temperature and
normalization become consistent with the {\em XMM} result, which
itself is consistent with previous results \citep[see, e.g.,][]{s02}.

We added a Gaussian feature centered at 6.4~keV to the continuum model
and determined the best fit width and flux of the line.  We find that
a wide, FWHM$=$2.64$\pm$0.14~keV, is allowed with a best-fit EW of
370$\pm$40~eV.  This line is comparable to that found by \citet{s02}
in the {\em RXTE} data of 4U~1543$-$624.  If the energy of the
Gaussian is allowed to vary, the Gaussian model becomes even wider,
FWHM$\approx$6~keV, with a central energy of 4~keV.  While it is
interesting to note that both the {\em XMM} and {\em RXTE} data shows
this similar feature, the large width and instability of the {\em XMM}
fits, leads us to believe that the feature is not a detection of Fe
line emission.

For the high-resolution measurements of the absorption edge depths, we
fit only the 2 RGS spectra using a power-law $+$ blackbody as the
continuum model.  We froze both the power-law index and the blackbody
temperature during the fits.  Again we used the variable absorption
model with O, Fe and Ne abundances set to zero, and included the
appropriate edge models for O, Fe, and Ne.  A Gaussian model was used
to model the atomic O absorption line and another was included to fit
a line feature at 23.36~\AA\/ which has been attributed to absorption
by O in metallic compounds \citep[i.e. oxides; e.g.,][]{scc+02} but is
also consistent with the \ion{O}{2} 1s-2p transition (see, Juett et
al. 2003).  We also found that the O edge could be better fit by
including 2 edge models.  The best-fit O edge wavelength is dependent
on the form of the O (atomic vs. metallic compounds) and the presence
of multiple forms of O would produce multiple edges.

The best-fit results are given in Tables 1 and 2 and shown in Figure
2.  From the fits of the {\em XMM} RGS spectra of 4U~1543$-$624, we
find $N_{\rm Ne}=(8.0\pm0.3)\times10^{17}$~cm$^{-2}$ and $N_{\rm
  Fe}=(4.7\pm0.6)\times10^{16}$~cm$^{-2}$.  For a single O edge, the
best-fit wavelength is 23.12$\pm$0.04~\AA\/ with a column density of
N$_{\rm O}=(14.8\pm0.2)\times10^{17}$ cm$^{-2}$.  Fitting with two O
edges yields wavelengths of 22.86$\pm$0.04~\AA\/ and
23.19$\pm$0.03~\AA\/ and column densities of
$(8.2\pm0.3)\times10^{17}$~cm$^{-2}$ and
$(6.5\pm0.3)\times10^{17}$~cm$^{-2}$.  The lower wavelength is
consistent with the predicted O edge for atomic O while the higher
wavelengths is expected from O bound in molecules.  We find a Ne/O
number ratio of 0.54$\pm$0.03.  For the atomic O absorption line, we
find a best-fit wavelength 23.515$\pm$0.006~\AA\/ and an EW of
0.0095$\pm$0.0011~\AA.  The other line has a wavelength of
23.371$\pm$0.009~\AA\/ and an EW of 0.0083$\pm$0.0014~\AA.  Only the
Fe-L column density is consistent between the {\em Chandra} and {\em
  XMM} observations.  The {\em XMM} spectral analysis shows an
increase in the O column density by a factor of 1.7, relative to the
{\em Chandra} result, and a reduction by 60\% in the Ne column
density.  While smaller than found in the {\em Chandra} spectrum, the
Ne/O ratio of 0.54 is still significantly larger than the ISM value of
0.18.

We also searched for line features in the {\em XMM} spectrum of
4U~1543$-$624 using the same procedure as for the {\em Chandra}
spectra.  Gaussian models with fixed FWHM$=$2000~km~s$^{-1}$ were fit
at each point in the wavelength range 6.2--35~\AA\/ for both RGS
spectra.  We place a 3$\sigma$ upper limit of 0.02~\AA\/ on the EW of
any line feature in the {\em XMM} spectrum of 4U~1543$-$624 in the
wavelength range 6--18~\AA.  At larger wavelengths, the EW limit
increases to 0.3~\AA\/ at 35~\AA.  These EW limits are much lower than
found in the 4U~1626$-$67 spectrum.  When the FWHM$=$5000~km~s$^{-1}$,
the EW limits are a factor of 2 greater.  As mentioned earlier, there
are definite residuals in the spectral fit of 4U~1543$-$624 (see,
Figure 5).  We attribute these residuals to instrumental calibration
uncertainty.

\begin{figure}
\epsscale{1.2}
\plotone{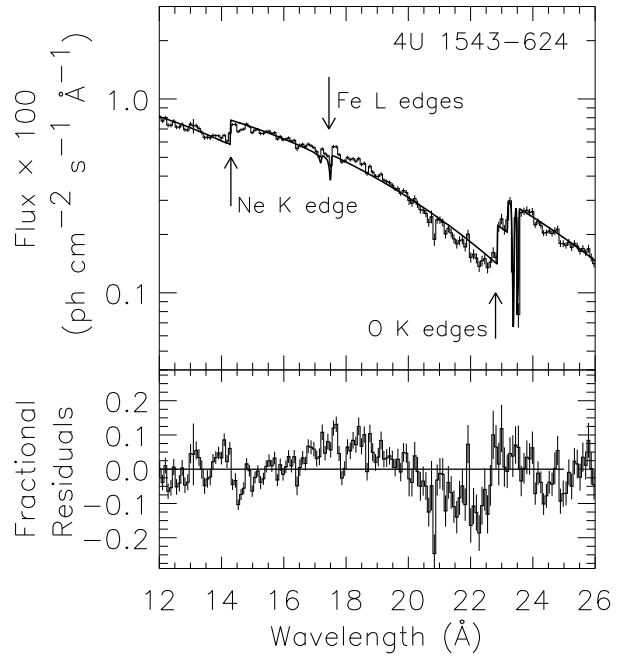}
\caption{(upper panel) {\em XMM} combined RGS first order spectrum of
4U~1543$-$624 with best-fit power-law $+$ blackbody model including
the O, Fe, and Ne edge models.  For clarity, the data is binned to
0.09~\AA.  (lower panel) Fractional residuals ([data$-$model]/model)
of the RGS first order spectral fit shown above.  The absorption line
at 23.5~\AA\/ is the expected atomic O line while the line at
23.37~\AA\/ is consistent with O absorption in molecular compounds.
There are no other features detected which are consistent with narrow
emission or absorption lines.  The arrows mark the Ne-K, Fe-L and O-K
absorption edges at 14.3, 17.5, and 23~\AA\/ respectively.}
\label{fig:5} 
\end{figure}

\subsection{Residual Features}
We now address the residuals found in our spectral analysis.  For the
{\em XMM} spectra of both sources, there are broad residual features
at the 10--20\% level.  These residuals are consistent with the
uncertainty in the calibration of the RGS\footnote{See document
XMM-SOC-CAL-TN-0030 available at
http://xmm.vilspa.esa.es/external/xmm\_sw\_cal/calib/index.shtml}.
All spectra show some level of structure around the neutral Ne-K edge
at 14.3~\AA.  For both instruments, the RGS and HETG, the
interpretation of this feature is complicated by instrumental features
at the same wavelength: the RGS has lost one of the CCDs that covered
the 10.5--14.0~\AA\/ wavelength range reducing the effective area in
this range, while the HETG nominal pointing places a chip gap in the
$-1$ MEG order at the Ne-K edge.

\begin{deluxetable*}{lcccc}
\tablecaption{Photoelectric Absorption Edge Measurements\tablenotemark{a}} 
\tablehead{ & \colhead{$\lambda$} & \colhead{Optical depth} & 
\colhead{$N_{\rm Z}$} & \colhead{$N_{\rm H}$} \\ 
\colhead{Edge} & \colhead{(\AA)} & \colhead{$\tau$} & 
\colhead{(10$^{17}$ cm$^{-2}$)} & \colhead{(10$^{21}$ cm$^{-2}$)}}
\startdata 
\multicolumn{5}{c}{2S~0918$-$549:{\em Chandra} result} \\ \tableline 
O K & 23.0$\pm$0.3 & 1.08$\pm$0.18 & 19$\pm$3 & 3.9$\pm$0.6 \\
Fe L\ion{}{3} & 17.49$\pm$0.02\tablenotemark{b} & 0.7$\pm$0.4 & 1.0$\pm$0.6 
& 3.7$\pm$2.2 \\
Ne K & 14.24$^{+0.2}_{-0.02}$ & 0.36$\pm$0.06 & 9.9$\pm$1.7 & 11.4$\pm$1.9 \\
\tableline
\multicolumn{5}{c}{2S~0918$-$549:{\em XMM} result} \\ \tableline 
O K & 22.99$\pm$0.08 & 1.11$\pm$0.04 & 19.6$\pm$0.7 & 4.00$\pm$0.14 \\
Fe L\ion{}{3} & 17.50$\pm$0.03\tablenotemark{b} & 0.6$\pm$0.2 & 0.08$\pm$0.03 
& 3.0$\pm$1.1 \\
Ne K & 14.40$\pm$0.03 & 0.33$\pm$0.02 & 9.1$\pm$0.6 & 10.4$\pm$0.6 \\
\tableline
\multicolumn{5}{c}{4U~1543$-$624:{\em Chandra} result} \\ \tableline 
O K & 22.72$^{+0.05}_{-0.03}$ & 0.48$\pm$0.08 & 8.5$\pm$1.4 & 1.7$\pm$0.3 \\
Fe L\ion{}{3} & 17.49$\pm$0.04\tablenotemark{b} & 0.21$\pm$0.14 & 0.3$\pm$0.2 
& 1.1$\pm$0.7 \\
Ne K & 14.330$\pm$0.017 & 0.45$\pm$0.04 & 12.4$\pm$1.1 & 14.2$\pm$1.3 \\ 
\tableline
\multicolumn{5}{c}{4U~1543$-$624:{\em XMM} result} \\ \tableline 
O K & 22.86$\pm$0.04 & 0.64$\pm$0.016 & 8.2$\pm$0.3 & 1.67$\pm$0.06 \\
O K & 23.19$\pm$0.03 & 0.368$\pm$0.016 & 6.5$\pm$0.3 & 1.32$\pm$0.06 \\
Fe L\ion{}{3} & 17.50$\pm$0.02\tablenotemark{b} & 0.33$\pm$0.04 & 
0.47$\pm$0.06 & 1.7$\pm$0.2 \\
Ne K & 14.286$^{+0.012}_{-0.05}$ & 0.291$\pm$0.010 & 8.0$\pm$0.3 & 
9.2$\pm$0.3 
\enddata 
\tablenotetext{a}{All errors are quoted at the 90\%-confidence level.
Atomic cross-sections taken from Henke et al. (1993) for O and Ne,
and Kortright \& Kim (2000) for Fe-L.  Interstellar abundances taken from 
\citet{wam00}.}
\tablenotetext{b}{We define the absorption edge energy for Fe as the
energy at which the edge structure reaches its minimum value.}
\end{deluxetable*}

While we do not claim a significant detection of any feature, the
similar structure seen in all spectra suggest that the simple
interpretation of a neutral Ne-K edge may not be entirely correct.
Below we detail some possible astrophysical origins, remembering that
instrumental effects are likely also important.

One explanation is that the feature is an emission line or a radiative
recombination continuum (RRC) emission feature. At 14.0--14.3~\AA,
emission from a number of different ionization states of Fe is
possible, but we would expect to see multiple features (e.g., as in
the spectrum of Cyg X-1; Schulz et al. 2002).  The \ion{O}{8} RRC
feature is also consistent with a 14.2~\AA\/ position.  Again though,
the lack of other \ion{O}{8} emission features, in particular the
\ion{O}{8} emission line at 18.97~\AA, makes this possibility less
probable.  

The \ion{O}{8} absorption edge is also located at 14.2~\AA\/ and the
{\em Chandra} spectrum of 4U~1543$-$624 (see, Figure 4) seems to
suggest that 2 edges, one at Ne-K and another at \ion{O}{8} may be a
plausible solution.  Finally, the feature might be related to the Ne
absorption edge.  Unlike other elements, Ne cannot be in molecular
forms which might cause structure at the K shell absorption edge.  We
can not rule out however unanticipated structure in the edge shape, in
particular if the Ne-rich absorbing material has some velocity, we
might expect to find a velocity shifted edge structure.  The
interpretation of this feature is hampered by the lack of
high-resolution atomic data on X-ray absorption edges.  As we acquire
more data on X-ray absorption edges, it will become possible to
determine if this structure is ``normal'' or instead suggests the
presence of more interesting phenomena.

\section{Discussion}
We have shown that the {\it Chandra}/HETGS and {\em XMM}/RGS spectra
of 2S~0918$-$549 and 4U~1543$-$624 are well fit by models that allow
for absorption columns of neutral Ne and O with abundance ratios
significantly different from the expected ISM ratio.  The {\em
Chandra} spectrum of 2S~0918$-$549 has a best fit Ne/O number ratio of
0.52$\pm$0.12, or roughly 3$\times$ the ISM value of 0.18.  The {\em
XMM} fit of 2S~0918$-$549 gives a consistent Ne/O number ratio of
0.46$\pm$0.03.  The best-fit model for the {\em Chandra} spectrum of
4U~1543$-$624 has a Ne/O number ratio of 1.5$\pm$0.3, while the
best-fit {\em XMM} model has a Ne/O number ratio of 0.54$\pm$0.03.
The unusual abundance ratios as well as the variability seen in the
observations of 4U~1543$-$624 lead us to conclude that there is
absorption local to these binaries and that the material is Ne
enriched.

From the $L_{\rm X}/L_{\rm opt}$ ratios of 2S~0918$-$549 and
4U~1543$-$624, we expect these binaries to have orbital periods
$\lesssim$$60$~min based on the empirical relationship determined by
van Paradijs \& McClintock (1994).  We searched the {\it Chandra\/}
lightcurves for orbital modulation, but found no modulations larger
than 1--2\%.  {\em XMM} provided a more sensitive search for orbital
modulations, but we found no signatures of orbital modulation at
fractional rms upper limits of 0.14--0.3\%.  Confirmation of such
short orbital periods would place 2S~0918$-$549 and 4U~1543$-$624 in
the class of ultracompact LMXBs (P$_{\rm orb}$$\lesssim$80 min).
Ultracompact binaries require H-depleted or degenerate dwarf
companions (Joss et al. 1978; Nelson et al. 1986).  Such companions
would be expected to have non-standard abundances compared to ISM
values.  The {\it Chandra\/} spectrum of 4U~1626$-$67 revealed
absorption edges of C, O, and Ne which are 5 times larger than would
be predicted given the hydrogen column density measured in the UV.
For 4U~1626$-$67, the measured local abundance ratios, if the excess
material is assumed to have originated around the binary, are
consistent with the expected abundances in the chemically fractionated
core of a C-O or O-Ne-Mg WD (Schulz et al. 2001).  The evidence for Ne
absorption in 2S~0918$-$549 and 4U~1543$-$624 hints at a similarity
between these sources and 4U~1626$-$67.  In addition, there is
evidence that the optical spectra of this group of sources is also
similar to 4U~1626$-$67 with no H or He lines detected, but with a
\ion{C}{3}/\ion{N}{3} emission line near 4640~\AA\/ (Wang \&
Chakrabarty 2003, in preparation).  Based on the similarity between
these sources and 4U~1626$-$67, we previously attributed the excess Ne
absorption in 2S~0918$-$549 and 4U~1543$-$624 to material local to the
sources, and suggested that these systems contained a Ne-rich
degenerate donor \citep{jpc01}.

The one assumption we have made in this analysis is that the
absorption is from neutral material only.  If there is a sizable
contribution to the absorption from material local to the binary, this
assumption is most likely not valid, since we would expect ionization
of the local material from the central source.  Accounting for
ionization of the material may explain the large Ne/O ratios observed.
We find Ne/O ratios larger than the 0.22 inferred for the local
absorption in the {\em Chandra} spectrum of 4U~1626$-$67
\citep{scm+01}.  Ne/O ratios for local absorption in our source will
be even higher than the measured total line-of-sight ratio once ISM
contributions are removed, since the O column from the ISM should be
larger than the Ne column. Unfortunately, we can not do this
calculation, as was possible for 4U~1626$-$67, since we do not know
the expected ISM contribution for 2S~0918$-$549 or 4U~1543$-$624.  If
the local material is affected by ionization, the effect will be
different for each element.  O will become ionized before Ne, leading
to an enhanced Ne/O ratio as measured by the neutral edges.  While it
is tempting to assume that ionization is the only cause for the
unusual abundances, we point out that if the local material was of
standard abundances, we would expect local absorption from higher Z
elements, like Mg and Si, which are not seen in the spectra of
2S~0918$-$549 or 4U~1543$-$624.  This leads us to conclude that
material must have some enhancement of Ne to show such strong
absorption.

One of the most interesting results is the difference in the Ne and O
column densities in the two observations of 4U~1543$-$624.  The {\em
XMM} results show a increase in the O column density with a decrease
in the Ne column density.  In addition, the {\em XMM} spectrum is both
softer and has a lower luminosity in the 2--10~keV band than the {\em
Chandra} spectrum, which suggests that the source was not in the same
state in the {\em Chandra} and {\em XMM} observations.  We suggest
that the difference in the 4U~1543$-$624 results provides support for
the local nature of the absorption.  Since the high-resolution results
of 2S~0918$-$549 are consistent not only in the measured Ne/O number
ratio but also in the spectral model in general, we are confident that
instrumental differences do not have a significant effect on the
spectral results.  It is interesting to note the the Ne/O ratios found
in the {\em Chandra} and {\em XMM} observations of 2S~0918$-$549 are
smaller than the inferred value from the {\em ASCA} result of
\citet{jpc01}, while at the same time the flux of 2S~0918$-$549 has
decreased by a factor of 10 since the {\em ASCA} observation.  It is
possible that the state changes alter the ionization structure of the
local material, which in turn changes the absorbing columns of neutral
Ne and O.  It will be difficult to separate the effect of unusual
abundances from ionization changes without more data on these sources.
We propose that multiple observations of 4U~1543$-$624 using
simultaneous high-resolution and broadband observations will probe the
connection between spectral state differences and the column density
variations.  It is important to note that ionization effects and the
ISM contribution must be understood before it is possible to determine
the intrinsic Ne/O abundance ratio of the local material, which can
then be used to place constraints on the composition of the companion.

Although we attribute the excess Ne absorption in 2S~0918$-$549 and
4U~1543$-$624 to local material in both of the binaries, we also
consider the possibility that the excess Ne is due to enhancements of
the ISM along the line of sight.  The best measurement of absorption
toward an X-ray binary, the {\it Chandra\/} spectral analysis of Cyg
X-1, shows columns of O and Ne that are consistent with standard ISM
abundances (Schulz et al. 2002).  To quantify the variations of the
ISM abundances, we have undertaken a study of the ISM using column
density measurements from the spectra of X-ray binaries (Juett et
al. 2003).  Initial results show that other LMXBs do not show the Ne/O
ratios measured for these sources.  At most, Ne/O number ratios are
only twice the expected ISM values.  This is still significantly
smaller than we find in 2S~0918$-$549 and 4U~1543$-$624, but does
suggest that there are some variations between lines-of-sight.  We
still favor the interpretation that there is absorbing material local
to the binaries, and that this is strengthened by the
observation-to-observation variations in the Ne/O number ratio of
4U~1543$-$624 which could not be due to the ISM.

Analysis of the {\it Chandra\/} and {\em XMM} spectra of 2S~0918$-$549
and 4U~1543$-$624 found no Ne or O lines like those seen in
4U~1626$-$67.  The absence of strong lines in the spectra of
2S~0918$-$549 and 4U~1543$-$624 demonstrates that their circumbinary
environments are different compared to 4U~1626$-$67.  One possible
reason for this difference may be that 4U~1626$-$67 is a pulsar with
$B$$=$$3\times10^{12}$~G (Orlandini et al. 1998), while 2S~0918$-$549
and 4U~1543$-$624 are likely to have weak magnetic fields
($\sim$10$^8$ G).  This undoubtedly results in a substantial
difference in the accretion flow geometry.  There is weak evidence for
a line-like feature at 6.4~keV in the {\em XMM} spectrum of
4U~1543$-$624.  This feature is very broad and could possibly be due
to a mis-modeling of the continuum emission rather than a Fe line.  In
addition, there is evidence for structure around the Ne edge in the
spectra of both sources.  It is likely that this is due to
instrumental effects but could also be due to structure in either
neutral Ne edges or possibly due to confusion with ionized O edges.

Recent {\it RXTE\/} and {\em BeppoSAX} observations of 2S~0918$-$549
found the first thermonuclear X-ray bursts from this source
\citep{jvh+01,cvz+02}.  Both bursts were short, with durations
$\lesssim$$100$~s.  These short bursts suggest that H and/or He is
undergoing unstable burning on the surface of the NS.  If the
companion has a large C abundance, we might expect to see a much
longer burst ($\approx$1 hr), like the ``superbursts'' seen in some
LMXBs, which have been attributed to thermonuclear burning of C on
neutron star surfaces (Cornelisse et al. 2000; Cumming \& Bildsten
2001; Strohmayer \& Brown 2001).  The more ordinary properties of the
X-ray bursts from 2S~0918$-$549 may indicate that the donor is not a
C-O dwarf as suggested for 4U~1626$-$67.  In this case, another
mechanism for Ne enhancement is needed.  On the other hand, it may
still be possible for a carbon accretor to show short X-ray bursts.
One possibility is that while the companion is H-deficient, there is
still a non-negligible H fraction that is accreted by the NS and is
then responsible for the Type I X-ray bursts \citep[e.g.,][]{nrj86}.
Alternatively, the heavy elements (C, O, and Ne) may undergo
spallation during accretion, leaving He and H nuclei which could then
undergo unstable thermonuclear burning in the usual way (Bildsten,
Salpeter, \& Wasserman 1992).  Spallation reactions would produce
$\gamma$-ray emission lines at 4.4 and 6.1 MeV.  Unfortunately, the
strength of these lines as calculated by Bildsten et al. (1992) is
below current observational detection limits.  For a 1~Ms observation
with {\em Integral}, the detection sensitivities are $\sim$1000 times
higher than the most optimistic line flux estimates.

\acknowledgements{We would like to thank the anonymous referee for his
comments which greatly improved this paper.  For help in determining
the optical position of 2S~0918$-$549, we thank Zhongxiang Wang.  We
are grateful to the HETGS instrument team at MIT for their help and
advice in analyzing these data.  We especially thank John Davis for
assistance in analyzing piled-up grating data, John Houck for his help
with and modifications of ISIS, and Herman Marshall for information on
the ACIS calibration status and the appropriate QE correction table.
We also acknowledge useful discussions with Lars Bildsten, Andrew
Cumming, Duncan Galloway, Peter Jonker, Eric Kuulkers, Jon Miller,
Mike Muno, Eric Pfahl, Dimitrios Psaltis, and Norbert Schulz.  This
work was supported in part by NASA under contracts NAS8-38249 and
NAS8-01129 and grant NAG5-9184.}

\end{document}